\documentclass[aps,prb,twocolumn,superscriptaddress]{revtex4}

\usepackage{graphicx}
\input{epsf}
\begin{document}

\author{R. Allub}
\thanks{Member of the Carrera del Investigador Cient\'{\i}fico del Consejo Nacional
de Investigaciones Cient\'{\i}ficas y t\'{e}cnicas (CONICET).}
\affiliation{Centro At\'{o}mico Bariloche, (8400) S. C. de Bariloche, Argentina.}

\title{Two-Impurity Anderson model in an Antiferromagnetic metal: zero-bandwidth limit.}

\begin{abstract}
We study the zero-bandwidth limit of the two-impurity Anderson model in an
antiferromagnetic (AF) metal. We calculate, for different values of the
model parameters, the lowest excitation energy, the magnetic correlation $<\mathbf{S}_{1}\mathbf{S}_{2}>$ between the impurities, and the magnetic
moment at each impurity site, as a function of the distance between the
impurities and the temperature. At zero temperature, in the region of
parameters corresponding to the Kondo regime of the impurities, we observe
an interesting competition between the AF gap and the Kondo physics of the
two impurities. When the impurities are close enough, the AF splitting
governs the physics of the system and the local moments of the impurities
are frozen, in a state with very strong ferromagnetic correlation between
the impurities and roughly independent of the distance. On the contrary,
when the impurities are sufficiently far apart and the AF gap is not too
large, the scenario of the Kondo physics take place: non-magnetic ground
state and the possibility of spin-flip excitation emerges and the
ferromagnetic $<\mathbf{S}_{1}\mathbf{S}_{2}>$ decreases as the distance
increases, but the complete decoupling of the impurities never occurs. In
adition, the presence of the AF gap gives a non-zero magnetic moment at each
impurity site, showing a non complete Kondo screening of the impurities in
the system. We observe that the residual magnetic moment decreases when the
distance between the impurities is increased.

\end{abstract}

\maketitle

%\pagebreak

\section{INTRODUCTION}

The behavior of spin correlations in heavy fermion systems is still not
completely understood. \cite{Lee} At high temperatures, heavy fermion
materials behave like a collection of individual local moments. When the
temperature goes down, correlations take place and the Kondo effect\cite%
{Kondo} can occur in this systems. This screening can quench the magnetic
interaction between local moments and the question is: how can spin
polarizations propagate to other local moments. A first approach toward the
understanding of this very interesting problem, the competition between
Kondo effect and Ruderman-Kittel-Kasuya-Yosida interaction (RKKY)\cite{RKKY}
has been studied in the simplified framework of the two-impurity Anderson%
\cite{Ander2} or Kondo\cite{Kon2} models. Recently, in the Kondo limit, a
systematic study of the ground states of two Anderson impurities has been
realized.\cite{Simonin} Both models take into account the conduction
electrons in a non-magnetic band. Nevertheless, many experiments in heavy
fermion systems show antiferromagnetic correlations or orderings at low
temperatures. For example, inelastic neutron scattering from the
antiferromagnetic heavy fermion system U$_{2}$Zn$_{17}$\cite{Bro} shows spin
fluctuations in this material. Also, UPt$_{3}$\cite{Aepp} and URu$_{2}$Si$_{2}$\cite{Broh} are both heavy fermion compounds where spin fluctuations
and antiferromagnetic correlations are present. This suggests that to
understand the anomalous properties of these materials, it is necessary to
study also the Kondo effect in presence of different kind of magnetic order
of itinerant electrons. Zhang and Yu\cite{Zha} considered a half-filled
anisotropic Kondo lattice model within a mean field theory and found 
a coexistence of antiferromagnetic long-range order and the Kondo singlet
state. Similar results are obtained by Capponi and Assaad\cite{Cap} using a
Quantum Monte Carlo algorithm. Recently, the single Kondo effect in an
antiferromagnetic metal was studied.\cite{Aji} This work shows that for a
general location of the impurity, the Kondo singularities still occur, but
the ground state has a partially unscreened moment. From the theoretical
point of view, a natural extension of this problem is to consider the case
of two-magnetic impurities in an antiferromagnetic metal. The study of a
pair of spin-1/2 impurities is a starting point for our understanding of a
lattice behavior in this kind of materials. The aim of this work is to
present a very simple approach to study how the competition between partial
quenching of the individual moments and their indirect interaction via the
antiferromagnetic conduction electrons take place. To this end, as a first
approach to the solution of this problem, we extend the zero band-width
(ZBW) limit approximation of the two-impurity Anderson model in a
paramagnetic metal\cite{Allub}, to include the AF conduction band.\cite{Aji}
Despite our simple approximation, the ZBW limit has been successfully
applied to explain qualitatively most of the experimental results in valence
fluctuating problems.\cite{angost} This limit also gives a good description
of the magnetic reentrance phenomena in superconductors with Kondo
impurities.\cite{reentr}Also this method was applied to explain transport
experiments on semiconductor quantum dots.\cite{Allu} An attractive feature
of the ZBW limit, is that all calculations can be realized exactly with a
minimum of numerical effort and the results are very satisfying, since they
reproduce results for properties found much more laboriously by other
techniques. For example, the most important results of our previous paper\cite{Allub} were obtained in Ref.\cite{Simonin} by means of variational
wave functions. Nevertheless, it is important
to recognize that the ZBW limit is oversimplified, specifically in not
containing any band structures and consequently we must expect to obtain a
cartoon of the real picture. In summary, motivated by the experimental results in the magnetic heavy fermion systems as mentioned above and by the previous successful 
theoretical work and following this track of thought, we employ the ZBW
limit to study this very interesting problem. In absence of more elaborated
theoretical solutions, this approach often gives results in a good
qualitative agreement with experimental data. 

We introduce the two-impurity Anderson Hamiltonian and set up the zero
band-width approximation to this problem in Section 2. Section 3 is devoted
to present the numerical results and discusses their physical implications.
Section 4 is devoted to conclusions.

\section{MODEL}

We start from the two-impurity Anderson Hamiltonian\cite{Anderso} in the
absence of direct hopping between impurities extended to include the
antiferromagnetism of the itinerant electrons: 
\begin{eqnarray}
H&=&\sum_{\mathbf{k},\sigma }\varepsilon _{(\mathbf{k)}}\text{ }c_{\mathbf{k}%
\sigma }^{\dagger }c_{\mathbf{k}\sigma }+ \sum_{\mathbf{k}}[\Gamma\text{ }%
(c_{\mathbf{k}\uparrow }^{\dagger }c_{\mathbf{k+Q}\uparrow }-c_{\mathbf{k}%
\downarrow }^{\dagger }c_{\mathbf{k+Q}\downarrow })+\text{H.c.}]  \nonumber
\\
&& +\varepsilon _{d}\sum_{\sigma ,j=1,2}\text{ }d_{j\sigma }^{\dagger
}d_{j\sigma }+U\sum_{j=1,2}\text{ }d_{j\uparrow }^{\dagger }d_{j\uparrow
}d_{j\downarrow }^{\dagger }d_{j\downarrow }  \nonumber \\
&& +\sum_{\mathbf{k},\sigma ,j=1,2}V_{kj}\text{ }(c_{\mathbf{k}\sigma
}^{\dagger }d_{j\sigma }\,+\text{H.c.}),
\end{eqnarray}

where $c_{\mathbf{k}\sigma }^{\dagger }$ ($c_{\mathbf{k}\sigma }$) creates
(destroys) an electron with momentum $\mathbf{k}$ and spin $\sigma $ in the
conduction band with energy $\varepsilon _{(\mathbf{k)}}$, and $d_{j\sigma
}^{\dagger }$ ($d_{j\sigma }$) creates (destroys) a localized electron with
spin $\sigma $ on the site $\mathbf{R}_{j}$ with energy $\varepsilon _{d}$.
Besides, $\Gamma $ is the AF gap, $\mathbf{Q}$ is the ordering wave-vector, $%
U$ is the localized-orbital Coulomb interaction, and $V_{kj}=Ve^{i\mathbf{%
k\cdot R}_{j}}$ where $V$ is the hybridization strength. For $\Gamma =0 $, $%
H $ reduces to the well-known two-impurity Anderson model\cite{Anderso}.
From the practical point of view, the ZBW approximation replaces the
structureless conduction bands by few states, located just at the Fermi
energy ($\varepsilon _{F} $); conceptually, this recognizes the fact that in
most experiments essentially only levels close to the Fermi energy are
relevant. As in the previous paper,\cite{Allub} we take here two different
vectors $\mathbf{k}$ ($\mathbf{k}_{1}$ and $\mathbf{k}_{2}$ with $\mathbf{k}_{1} \neq \mathbf{k}_{2}$ and $|\mathbf{k}_{1}|=|\mathbf{k}_{2}|=|\mathbf{k}_{F}|$, with $\mathbf{k}_{F} $ the Fermi momentum) as a minimal model to
compensate the two localized spins at the impurities sites. The model should
lead to two independent Anderson problems when the impurities are
sufficiently far apart and $\Gamma=0$. Accordingly, the original Hamiltonian
of Eq. (1) reduces to 
\begin{eqnarray}
H_{ZBW}&=&\varepsilon _{F}\sum_{\sigma }\text{ }(c_{k_{1}\sigma }^{\dagger
}c_{k_{1}\sigma }+c_{k_{2}\sigma }^{\dagger }c_{k_{2}\sigma })  \nonumber \\
&& +\Gamma\text{ }[(c_{k_{1}\uparrow }^{\dagger }c_{k_{2}\uparrow
}-c_{k_{1}\downarrow }^{\dagger }c_{k_{2}\downarrow })+\text{H.c.}] 
\nonumber \\
&& +\varepsilon _{d}\sum_{\sigma ,j=1,2}\text{ }d_{j\sigma }^{\dagger
}d_{j\sigma }+U\sum_{j=1,2}\text{ }d_{j\uparrow }^{\dagger }d_{j\uparrow
}d_{j\downarrow }^{\dagger }d_{j\downarrow }  \nonumber \\
&& +V\sum_{\sigma }(e^{i\phi _{1}}\text{ }c_{k_{1}\sigma }^{\dagger
}d_{1\sigma }\,+e^{i\phi _{1}^{\prime }}\text{ }c_{k_{2}\sigma }^{\dagger
}d_{1\sigma }  \nonumber \\
&&+e^{i\phi _{2}^{\prime }}\text{ }c_{k_{1}\sigma }^{\dagger }d_{2\sigma
}\,+e^{i\phi _{2}}\text{ } c_{k_{2}\sigma }^{\dagger }d_{2\sigma }+\text{H.c.%
}),
\end{eqnarray}

with $k_{1}=|\mathbf{k}_{1}|=|\mathbf{k}_{F}|$, $k_{2}=|\mathbf{k}_{2}|=|%
\mathbf{k}_{1}+\mathbf{Q}|=|\mathbf{k}_{F}|$, $\phi _{1}=\mathbf{k}_{1}\mathbf{\cdot R}_{1}=\mathbf{k}_{1}\mathbf{\cdot (R}_{2}+\mathbf{r}) $, $%
\phi _{1}^{\prime }=\mathbf{k}_{2}\mathbf{\cdot R}_{1}$,$\phi _{2}^{\prime }=\mathbf{k}_{1}\mathbf{\cdot R}_{2} $, and $\phi _{2}=\mathbf{k}_{2}\mathbf{%
\cdot R}_{2}=\mathbf{k}_{2}\mathbf{\cdot (R}_{1}-\mathbf{r}) $, where $\mathbf{r=R_{1}-R}_{2}$ is the distance between impurities. We can rewrite
the Eq. (2) in terms of $\mathbf{r}$, to this end we define $c_{1\sigma
}^{\dagger }=$ $e^{i\phi _{2}^{\prime }}c_{k_{1}\sigma }^{\dagger }$ and $%
c_{2\sigma }^{\dagger }=$ $e^{i\phi _{1}^{\prime }}c_{k_{2}\sigma }^{\dagger
}$:

\begin{eqnarray}
H_{ZBW} &=&\varepsilon _{F}\sum_{\sigma }\text{ }(c_{1\sigma }^{\dagger
}c_{1\sigma }+c_{2\sigma }^{\dagger }c_{2\sigma })  \nonumber \\
&& + \Gamma\text{ }[e^{-i(\mathbf{k}_{1}\mathbf{\cdot R_{2}}-\mathbf{k}_{2}%
\mathbf{\cdot R_{1}}) }(c_{1\uparrow }^{\dagger }c_{2\uparrow
}-c_{1\downarrow }^{\dagger }c_{2\downarrow })+\text{H.c.}]  \nonumber \\
&& +\varepsilon _{d}\sum_{\sigma ,j=1,2}\text{ }d_{j\sigma }^{\dagger
}d_{j\sigma }+U\sum_{j=1,2}\text{ }d_{j\uparrow }^{\dagger }d_{j\uparrow
}d_{j\downarrow }^{\dagger }d_{j\downarrow }  \nonumber \\
&& +V\sum_{\sigma }(e^{i\mathbf{k}_{1}\mathbf{\cdot r} }\text{ }c_{1\sigma
}^{\dagger }d_{1\sigma }\,+c_{2\sigma }^{\dagger }d_{1\sigma }  \nonumber \\
&& +c_{1\sigma }^{\dagger }d_{2\sigma }\,+e^{-i\mathbf{k}_{2}\mathbf{\cdot r}
}\text{ }c_{2\sigma }^{\dagger }d_{2\sigma }+\text{H.c.}),
\end{eqnarray}
To solve the model Hamiltonian with a minimal number of parameters we take $%
\mathbf{R}_{1}=\mathbf{r}/2$ and $\mathbf{R}_{2}=-\mathbf{r}/2$, and we
define $\phi =\mathbf{k}_{1}\mathbf{\cdot r}$ and $\phi ^{\prime }=\mathbf{Q}%
\mathbf{\cdot r}$. Then, we rewrite Eq.(3) as 
\begin{eqnarray}
H_{ZBW} &=&\varepsilon _{F}\sum_{\sigma }\text{ }(c_{1\sigma }^{\dagger
}c_{1\sigma }+c_{2\sigma }^{\dagger }c_{2\sigma })  \nonumber \\
&& + \Gamma\text{ }[e^{i(\phi + \phi ^{\prime }/2)}(c_{1\uparrow }^{\dagger
}c_{2\uparrow }-c_{1\downarrow }^{\dagger }c_{2\downarrow })+\text{H.c.}] 
\nonumber \\
&& +\varepsilon _{d}\sum_{\sigma ,j=1,2}\text{ }d_{j\sigma }^{\dagger
}d_{j\sigma }+U\sum_{j=1,2}\text{ }d_{j\uparrow }^{\dagger }d_{j\uparrow
}d_{j\downarrow }^{\dagger }d_{j\downarrow }  \nonumber \\
&& +V\sum_{\sigma }(e^{i\phi}\text{ }c_{1\sigma }^{\dagger }d_{1\sigma
}\,+c_{2\sigma }^{\dagger }d_{1\sigma }  \nonumber \\
&& +c_{1\sigma }^{\dagger }d_{2\sigma }\,+e^{-i(\phi + \phi ^{\prime })}%
\text{ }c_{2\sigma }^{\dagger }d_{2\sigma }+\text{H.c.}),
\end{eqnarray}
where we use $\mathbf{k}_{2}\mathbf{\cdot r}=(\mathbf{k}_{1}+\mathbf{Q})%
\mathbf{\cdot r}=(\phi + \phi ^{\prime })$ and $\mathbf{k}_{1}\mathbf{\cdot
R_{2}}-\mathbf{k}_{2}\mathbf{\cdot R_{1}}=-[\mathbf{k}_{1}\mathbf{\cdot r}+(%
\mathbf{k}_{1}+\mathbf{{Q})\cdot r]/2=-(\phi + \phi ^{\prime }/2)}$. For $%
r\rightarrow 0$, $\phi \rightarrow 0 $ and $\phi ^{\prime }\rightarrow 0$,
and we can write the model Hamiltonian as $H_{ZBW}=H_{0}+H_{1}$, with

\begin{eqnarray}
H_{0} &=&\sum_{\sigma }(\varepsilon _{F}+S_{\sigma }\Gamma)\text{ }\alpha
_{1\sigma }^{\dagger }\alpha _{1\sigma }+\varepsilon _{d}\sum_{\sigma ,j=1,2}%
\text{ } d_{j\sigma }^{\dagger }d_{j\sigma }  \nonumber \\
&& +U\sum_{j=1,2}\text{ }d_{j\uparrow }^{\dagger }d_{j\uparrow
}d_{j\downarrow }^{\dagger }d_{j\downarrow }  \nonumber \\
&& +\sqrt{2}V\sum_{\sigma }(\alpha _{1\sigma }^{\dagger }d_{1\sigma
}\,+\alpha _{1\sigma }^{\dagger }d_{2\sigma }+\text{H.c.})
\end{eqnarray}

and

\begin{equation}
H_{1}=\sum_{\sigma }(\varepsilon _{F}-S_{\sigma }\Gamma)\text{ }\alpha
_{2\sigma }^{\dagger }\alpha _{2\sigma },
\end{equation}

where we define $S_{\sigma }=+$ $(-)$ for the spin $\sigma =\uparrow $ $%
(\downarrow )$, $\alpha _{1\sigma }^{\dagger }=(c_{1\sigma }^{\dagger
}+c_{2\sigma }^{\dagger })/\sqrt{2}$ and $\alpha _{2\sigma }^{\dagger
}=(c_{1\sigma }^{\dagger }-c_{2\sigma }^{\dagger })/\sqrt{2}$. This is the
limit to the case in which the impurities are close together and we see that
the hybridization with the conduction band electrons reduces to only one
orbital ($\alpha _{1\sigma }$ )and favors the ferromagnetic coupling between
the impurities. So that, the $H_{ZBW}$ reduces to an effective simplified
zero band-width Hamiltonian ($H_{0}$) plus a diagonal term ($H_{1}$)
disconnected from it. In consequence, the mathematical problem reduces to
solve $H_{0}.$ For $\phi \neq 0$ and $\phi ^{\prime }=\pm \pi ,\pm 3\pi
,...,\pm (2n+1)\pi $ the hybridization term reduces to the case in which two
orthogonal band states are coupled each to a different impurity (i.e. $%
V\sum_{\sigma }[(e^{i\phi }c_{1\sigma }^{\dagger }\,+c_{2\sigma }^{\dagger
})d_{1\sigma }+(c_{1\sigma }^{\dagger }\,-e^{-i\phi }c_{2\sigma }^{\dagger
})d_{2\sigma }\,+$ H.c.$]=\sqrt{2}V\sum_{\sigma }[\gamma _{1\sigma
}^{\dagger }d_{1\sigma }+\gamma _{2\sigma }^{\dagger }d_{2\sigma }\,+$ H.c.$]
$, where $\gamma _{1\sigma }^{\dagger }=(e^{i\phi }c_{1\sigma }^{\dagger
}+c_{2\sigma }^{\dagger })/\sqrt{2}$ and $\gamma _{2\sigma }^{\dagger
}=(c_{1\sigma }^{\dagger }-e^{-i\phi }c_{2\sigma }^{\dagger })/\sqrt{2}$).
For $\Gamma =0$, this is the limit of the model when the impurities are
sufficiently far apart and the ZBW Hamiltonian reduces to two independent
Anderson problems. For $\Gamma \neq 0$ and $\phi ^{\prime }=\pm \pi $ the
antiferromagnetic term reduces to $\Gamma \text{ }[e^{i(\phi +\phi ^{\prime
}/2)}(c_{1\uparrow }^{\dagger }c_{2\uparrow }-c_{1\downarrow }^{\dagger
}c_{2\downarrow })+\text{H.c.}]=\Gamma \text{ }[\mp ie^{-i\phi }(\gamma
_{1\uparrow }^{\dagger }\gamma _{2\uparrow }-\gamma _{1\downarrow }^{\dagger
}\gamma _{2\downarrow })+\text{H.c.}]$. Therefore, for any value of $\phi $,
we can see that the impurities are always correlated due to the
antiferromagnetic order of the itinerant electrons. To study the interplay
between the hybridization and the antiferromagnetic order in this simple
theoretical picture we take hereafter the ordering wave-vector $\mathbf{Q}=-2%
\mathbf{k}_{1}$ ($\phi ^{\prime }=-2\phi $), no different physical results
are obtained with other values. So that, Eq. (4) reduces to 
\begin{eqnarray}
H_{ZBW} &=&\varepsilon _{F}\sum_{\sigma }\text{ }(c_{1\sigma }^{\dagger
}c_{1\sigma }+c_{2\sigma }^{\dagger }c_{2\sigma })  \nonumber \\
&&+\Gamma \text{ }[(c_{1\uparrow }^{\dagger }c_{2\uparrow }-c_{1\downarrow
}^{\dagger }c_{2\downarrow })+\text{H.c.}]  \nonumber \\
&&+\varepsilon _{d}\sum_{\sigma ,j=1,2}\text{ }d_{j\sigma }^{\dagger
}d_{j\sigma }+U\sum_{j=1,2}\text{ }d_{j\uparrow }^{\dagger }d_{j\uparrow
}d_{j\downarrow }^{\dagger }d_{j\downarrow }  \nonumber \\
&&+V\sum_{\sigma }(e^{i\phi }\text{ }c_{1\sigma }^{\dagger }d_{1\sigma
}\,+c_{2\sigma }^{\dagger }d_{1\sigma }  \nonumber \\
&&+c_{1\sigma }^{\dagger }d_{2\sigma }\,+e^{i\phi }\text{ }c_{2\sigma
}^{\dagger }d_{2\sigma }+\text{H.c.}).
\end{eqnarray}%
For $\phi =0$ Eq.(7) gives $H_{ZBW}=H_{0}+H_{1}$. For $\phi =\pi /2$ we have 
$\gamma _{1\sigma }^{\dagger }=(ic_{1\sigma }^{\dagger }+c_{2\sigma
}^{\dagger })/\sqrt{2}$ and $\gamma _{2\sigma }^{\dagger }=(c_{1\sigma
}^{\dagger }+ic_{2\sigma }^{\dagger })/\sqrt{2}$ and we can rewrite $H_{ZBW}$
in terms of two independent Anderson Hamiltonians plus a coupling term: $%
H_{ZBW}=H_{A1}+H_{A2}+\Gamma \text{ }[(\gamma _{1\uparrow }^{\dagger }\gamma
_{2\uparrow }-\gamma _{1\downarrow }^{\dagger }\gamma _{2\downarrow })+\text{%
H.c.}]$, where we define

\begin{eqnarray}
H_{Aj}&=&\varepsilon _{F}\sum_{\sigma }\text{ }\gamma _{j\sigma }^{\dagger
}\gamma _{j\sigma }+\varepsilon _{d}\sum_{\sigma }\text{ }d_{j\sigma
}^{\dagger }d_{j\sigma }+U\text{ }d_{j\uparrow }^{\dagger }d_{j\uparrow
}d_{j\downarrow }^{\dagger }d_{j\downarrow }  \nonumber \\
&& +\sqrt{2}V\sum_{\sigma }(\gamma _{j\sigma }^{\dagger }d_{j\sigma }\,+%
\text{H.c.}).
\end{eqnarray}
  
\section{RESULTS AND DISCUSSION}

\begin{figure}[ht]
\begin{center}
\includegraphics[width=6cm]{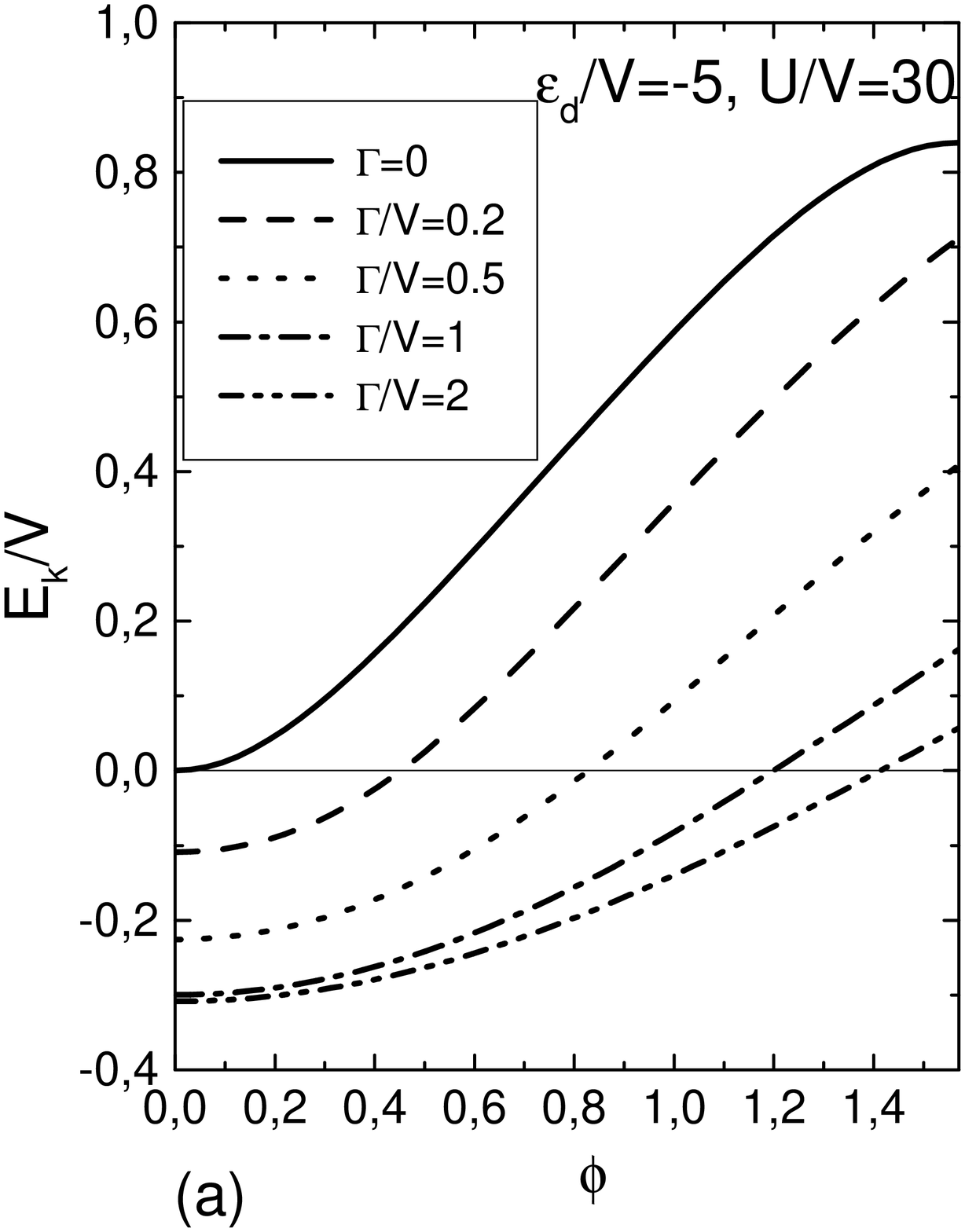} \includegraphics[width=6cm]{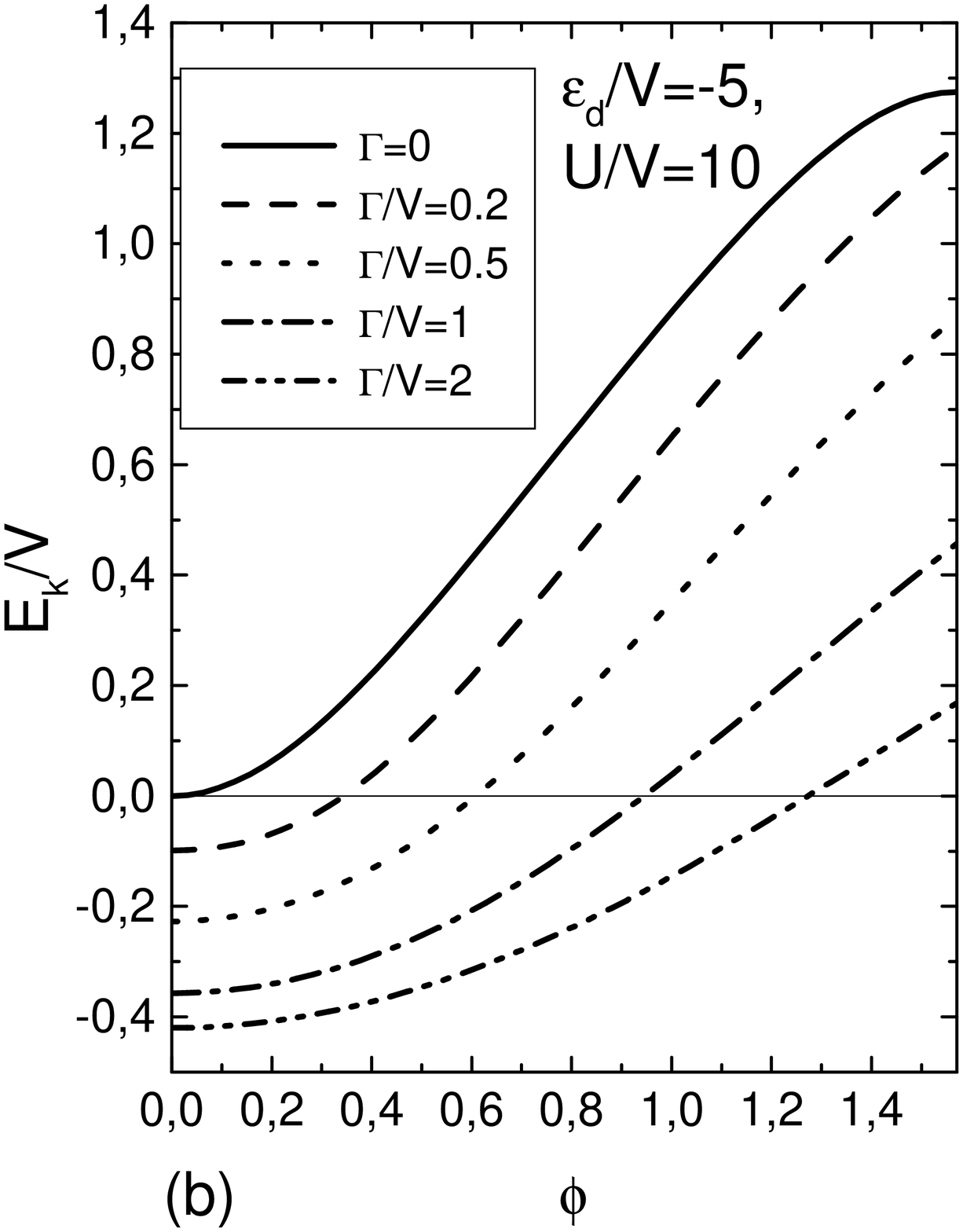} 
\includegraphics[width=6cm]{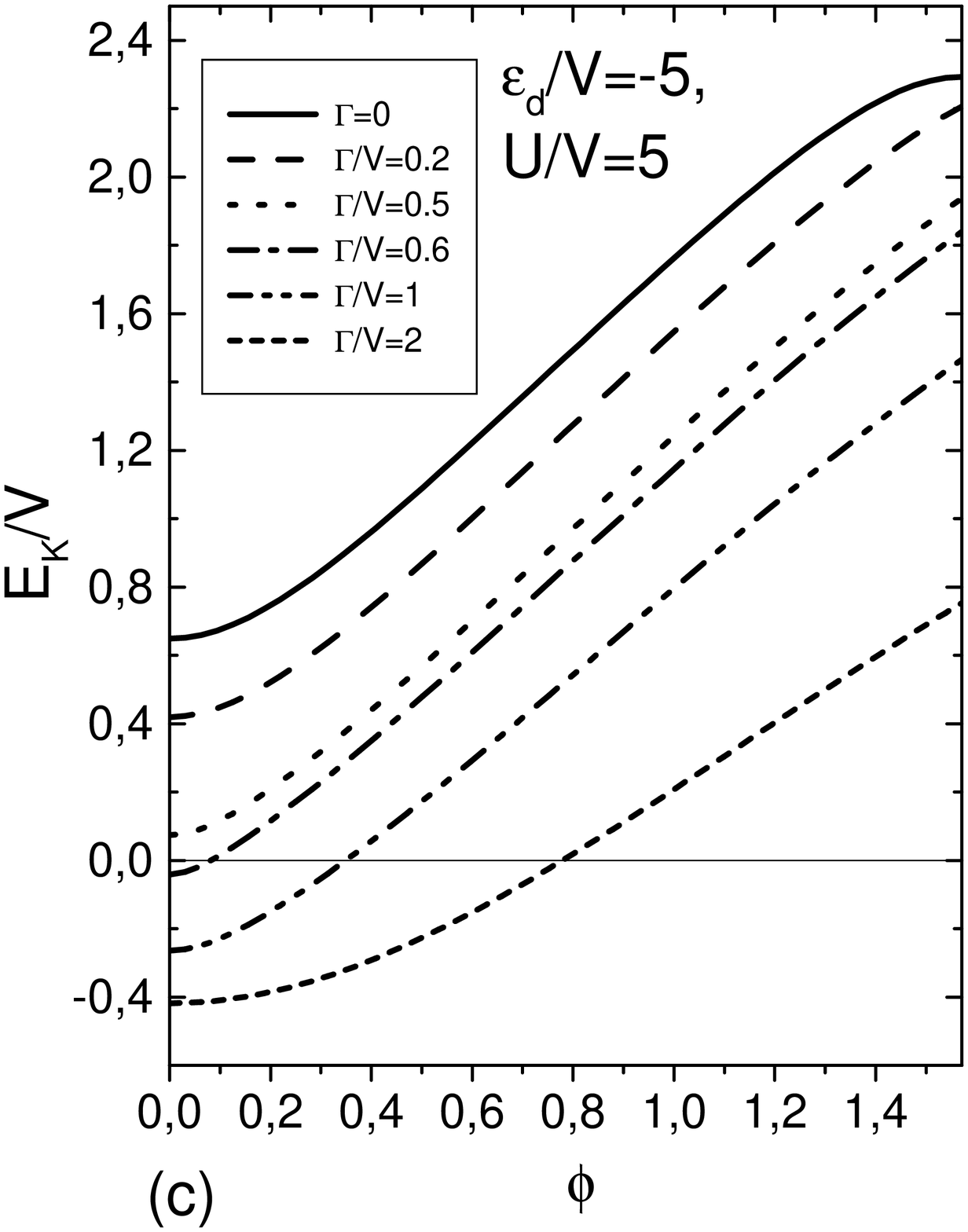}
\end{center}
\caption{The energy difference $E_{K}=(\protect\lambda _{S_{z}=+1}-\protect\lambda _{S_{z}=0})$ as a function of $\protect\phi $, for $\protect\varepsilon _{d}/V=-5$, five different values of $\Gamma/V= $ 0, 0.2, 0.5,
1, and 2, and three different values of $U/V= $ 30 (a), 10 (b), and 5 (c)}
\label{fig1}
\end{figure}

The magnetic correlations between the impurities given by the model
Hamiltonian (Eq.7) can be obtained from the four-particle states (this is
the most relevant Hilbert space in relation to the two Anderson problems
discussed here) or from the grand canonical ensemble adjusting the chemical
potential in such a way that the mean total number of particles is always
four. There is little numerical difference between these alternative
calculations.\cite{angost} So that, in all the numerical results presented
below, we use the four-particle states ($N=4 $). For this case, the full
Hamiltonian matrix is 70$\times $70. Nevertheless, the solution of the
problem reduces to the diagonalization of two 16$\times $16 matrices (for $%
S_{z}=\pm1$) and 36$\times $36 matrix (for $S_{z}=0$) as the full Hilbert
space is block diagonalized, with each block corresponding to a given $S_{z}$%
-component. For $S_{z}=\pm2$, the model gives two degenerate eigenvalues ($%
\lambda _{2}=2(\varepsilon _{F}+\varepsilon _{d})$). To obtain the numerical
results we take the Fermi energy $\varepsilon _{F}=0$ and $V$ as the unit of
energy. Therefore, the model is completely characterized by $\varepsilon
_{d} $, $U$, $\Gamma $ and the parameter $\phi $ as a measure of the
distance between the impurities. We start by presenting in Fig.~\ref{fig1}
the energy difference of the two lowest energy levels $E_{K}=(\lambda
_{S_{z}=+1}-\lambda _{S_{z}=0})$ as a function of $\phi $, for $\varepsilon
_{d}/V=-5$, five different values of $\Gamma/V= $ 0, 0.2, 0.5, 1, and 2, and
three different values of $U/V= $ 30, 10, and 5 (Fig.~\ref{fig1}(a), (b),
and (c)respectively) ranging from the Kondo limit ($U>>|\varepsilon
_{d}-\varepsilon _{F}|$) to the intermediate valence (I.V.) regime ($U
\sim|\varepsilon _{d}-\varepsilon _{F}|$). We can see that $E_{K} $ always
decreases when $\phi $ decreases and also $E_{K} $ decreases when $\Gamma $
is increases. In the Kondo limit (Fig.~\ref{fig1}(a)and (b)), for any value $%
\Gamma/V \neq 0 $, $E_{K}< 0 $ for $\phi=0 $ and $E_{K}> 0 $ for $\phi =
\pi/2 $. Therefore, for $\Gamma \neq0$, there is a particular value $\phi=
\phi_{c}$, with $0<\phi_{c}<\pi/2 $, where $E_{K}=0 $. For $\phi< \phi_{c} $
the ground state properties correspond to $S_{z}=+1 $ state (if $\Gamma < 0 $%
, $S_{z}=-1 $) and for $\phi> \phi_{c} $ the $S_{z}=0 $ ground state
properties take place. In the I.V. regime (Fig.~\ref{fig1}(c)), for a given
value of $U /|\varepsilon _{d}-\varepsilon _{F}|$, the existence or not of $%
\phi_{c} $ depend on the value of $\Gamma/V $. When $U/V $ reduces, large
values of $\Gamma/V $ are needed to obtain $S_{z}=+1 $ ground state at $%
\phi=0 $.

To analyze these results, we consider first the limit of $\phi=\pi /2$,
where we have the Hamiltonian $H_{ZBW}=H_{A1}+H_{A2}+\Gamma\text{ }%
[(\gamma_{1\uparrow }^{\dagger }\gamma_{2\uparrow }-\gamma_{1\downarrow
}^{\dagger }\gamma_{2\downarrow })+\text{H.c.}]$. For any value of $\Gamma $%
, the solution gives $S_{z}=0$ ground state. Therefore, in this limit we
have always $E_{K} > 0 $ (see Fig.~\ref{fig1}). It is easy to show this fact
in the Kondo limit of each impurity ($U/|\Delta |\rightarrow \infty $, $V/|\Delta |<<1$, with $\Delta =(\varepsilon _{d}-\varepsilon _{F})/2$). In
this limit, the 36$\times $36 matrix can be simplified to obtain,
approximately, the ground state energy of $H_{ZBW}$ and the corresponding
eigenvector ($a_{1},a_{2}$, and $a_{3} $) from the lowest eigenvalue $\lambda _{4,0}$ of the 3x3 matrix given by

\begin{equation}
\hspace{10mm}\left| 
\begin{array}{ccc}
2\varepsilon _{d}+2J & \Gamma & 0 \\ 
\Gamma & 2\varepsilon _{d}+J/2 & -\sqrt{3}\Gamma \\ 
0 & -\sqrt{3}\Gamma & 2\varepsilon _{d}%
\end{array}
\right| ,
\end{equation}
with $J=-2V^{2}/|\Delta |$. To this case the ground state reads ($%
|N,S_{z}>_{\phi}$): $|4,0>_{\pi /2}=\sum_{i}a_{i}|gi> $, where we define $%
|g1>=\frac{1}{2}[-\gamma_{1\uparrow }^{\dagger}\gamma_{2\uparrow}^{\dagger
}|->-\gamma_{1\downarrow }^{\dagger}\gamma_{2\downarrow}^{\dagger }|+>+\frac{%
1}{\sqrt{2}}(\gamma_{1\uparrow }^{\dagger}\gamma_{2\downarrow}^{\dagger
}+\gamma_{1\downarrow }^{\dagger}\gamma_{2\uparrow}^{\dagger })|\Theta >-%
\frac{1}{\sqrt{2}}(\gamma_{1\uparrow
}^{\dagger}\gamma_{2\downarrow}^{\dagger }-\gamma_{1\downarrow
}^{\dagger}\gamma_{2\uparrow}^{\dagger })|\Omega >]$, $|g2>=\frac{1}{\sqrt{2}%
}(\gamma_{2\uparrow }^{\dagger}\gamma_{2\downarrow}^{\dagger
}-\gamma_{1\uparrow }^{\dagger}\gamma_{1\downarrow}^{\dagger })|\Theta >$,
and $|g3>= \frac{(-1)}{2\sqrt{3}}[\gamma_{1\uparrow
}^{\dagger}\gamma_{2\uparrow}^{\dagger }|->+\gamma_{1\downarrow
}^{\dagger}\gamma_{2\downarrow}^{\dagger }|+>+3\frac{1}{\sqrt{2}}%
(\gamma_{1\uparrow }^{\dagger}\gamma_{2\downarrow}^{\dagger
}+\gamma_{1\downarrow }^{\dagger}\gamma_{2\uparrow}^{\dagger })|\Theta >+%
\frac{1}{\sqrt{2}}(\gamma_{1\uparrow
}^{\dagger}\gamma_{2\downarrow}^{\dagger }-\gamma_{1\downarrow
}^{\dagger}\gamma_{2\uparrow}^{\dagger })|\Omega >]$, with $|+>=d_{1\uparrow
}^{\dagger }d_{2\uparrow }^{\dagger }|0> $, $|->=d_{1\downarrow }^{\dagger
}d_{2\downarrow }^{\dagger }|0>$, $|\Theta >=\frac{1}{\sqrt{2}}(d_{1\uparrow
}^{\dagger }d_{2\downarrow }^{\dagger }+d_{1\downarrow }^{\dagger
}d_{2\uparrow }^{\dagger })|0>$, and $|\Omega >=\frac{1}{\sqrt{2}}%
(d_{1\uparrow }^{\dagger }d_{2\downarrow }^{\dagger }-d_{1\downarrow
}^{\dagger }d_{2\uparrow }^{\dagger })|0>$. Note that, $|g1>$ is the product
of the two Kondo (one for each impurity) singlet states: $\frac{1}{\sqrt{2}}%
(\gamma_{1\uparrow }^{\dagger }d_{1\downarrow }^{\dagger
}-\gamma_{1\downarrow }^{\dagger }d_{1\uparrow}^{\dagger })*\frac{1}{\sqrt{2}%
}(\gamma_{2\uparrow }^{\dagger }d_{2\downarrow }^{\dagger
}-\gamma_{2\downarrow }^{\dagger }d_{2\uparrow}^{\dagger })|0>$. The
explicit form of $|4,0>_{\pi /2}$ shows clearly the important contribution
of the ferromagnetic correlations between the impurities in the ground
state. Solving the cubic equation we obtain the ground state energy. We can
write, approximately, two limiting cases: $\lambda _{4,0}\simeq 2\varepsilon
_{d}-2\Gamma+\frac{J}{2}-\frac{3}{16}\frac{J^{2}}{\Gamma}$ for $\Gamma\gg
|J| $ and $\lambda _{4,0}\simeq 2\varepsilon _{d}+2J-\frac{2}{3}\frac{%
\Gamma^{2}}{|J|}$ for $|J|\gg \Gamma$. In a similar manner, the
simplification of the 16$\times $16 matrices (for $S_{z}=\pm1$) in the Kondo
limit, allow us to obtain the first excited energy level from the 3$\times $%
3 matrix given by 
\begin{equation}
\hspace{10mm}\left| 
\begin{array}{ccc}
2\varepsilon _{d}+J & \sqrt{2}\Gamma & 0 \\ 
\sqrt{2}\Gamma & 2\varepsilon _{d}+J/2 & \sqrt{2}\Gamma \\ 
0 & \sqrt{2}\Gamma & 2\varepsilon _{d}%
\end{array}
\right| .
\end{equation}
The lowest eigenvalue gives $\lambda _{4,1}=2\varepsilon _{d}+\frac{J}{2}-%
\sqrt{(\frac{J}{2}) ^{2}+4\Gamma^{2}}$. Therefore, for $\Gamma > 0$ and $%
\phi=\pi /2$ we obtain, approximately, $E_{K}\simeq \frac{J^{2}}{8\Gamma}$
for $\Gamma\gg |J|$ and $E_{K}\simeq -J-\frac{10}{3}\frac{\Gamma^{2}}{|J|} $
for $|J|\gg \Gamma$. For $\Gamma=0$ (solid lines in Fig.~\ref{fig1}(a) and
(b)), the problem reduces to solve the one impurity problem ($H_{Aj} $) and
we have obtained\cite{Allub} $E_{K}=\Delta +R_{0} $, with $R_{0}=\sqrt{%
\Delta^{2}+4V^{2}}$.

In the opposite limit, for $\phi =0$, the model gives $H_{ZBW}=H_{0}+H_{1}$
and we find that two different ground states are possible:

A) For large Coulomb repulsion ($|\varepsilon _{d}-\varepsilon _{F}| << U$),
the three-particle states and $S_{z}=+1/2$ (for $\Gamma >0 $), gives the
ground state energy of $H_{0}$ and the corresponding eigenvector ($%
b_{1},b_{2},b_{3},b_{4}$, and $b_{5}$) can be obtained easily from the
lowest eigenvalue $\lambda _{3,+1/2}$ of the 5x5 matrix given by

\begin{equation}
\hspace{10mm}\left\vert 
\begin{array}{ccccc}
\varepsilon _{1}-\Gamma  & 0 & 2V & 0 & -2V \\ 
0 & \varepsilon _{1}+\Gamma  & -\sqrt{2}V & 0 & \sqrt{2}V \\ 
2V & -\sqrt{2}V & \varepsilon _{2} & -\sqrt{2}V & 0 \\ 
0 & 0 & -\sqrt{2}V & \varepsilon _{3}+\Gamma  & \sqrt{2}V \\ 
-2V & \sqrt{2}V & 0 & \sqrt{2}V & \varepsilon _{4}%
\end{array}%
\right\vert ,
\end{equation}%
where $\varepsilon _{1}=2\varepsilon _{d}+\varepsilon _{F}$, $\varepsilon
_{2}=\varepsilon _{d}+2\varepsilon _{F}$, $\varepsilon _{3}=2\varepsilon
_{d}+\varepsilon _{F}+U$, and $\varepsilon _{4}=3\varepsilon _{d}+U$. So
that, the ground state of $H_{0}$ can be written as: $|3,+1/2>_{0}=%
\sum_{i}b_{i}|hi>$, where we define $|h1>=\alpha _{1\downarrow }^{\dagger
}|+>$, $|h2>=\alpha _{1\uparrow }^{\dagger }|\Theta >$, $|h3>=\frac{1}{\sqrt{%
2}}\alpha _{1\uparrow }^{\dagger }\alpha _{1\downarrow }^{\dagger
}(d_{1\uparrow }^{\dagger }-d_{2\uparrow }^{\dagger })|0>$ , $|h4>=\frac{1}{%
\sqrt{2}}\alpha _{1\uparrow }^{\dagger }(d_{1\uparrow }^{\dagger
}d_{1\downarrow }^{\dagger }-d_{2\uparrow }^{\dagger }d_{2\downarrow
}^{\dagger })|0>$, and $|h5>=\frac{1}{\sqrt{2}}(d_{1\uparrow }^{\dagger
}d_{1\downarrow }^{\dagger }d_{2\uparrow }^{\dagger }-d_{1\uparrow
}^{\dagger }d_{2\uparrow }^{\dagger }d_{2\downarrow }^{\dagger })|0>.$ In a
similar manner, we obtain the first excited state as $|3,-1/2>_{0}=%
\sum_{i}b_{i}^{\prime }|h^{\prime }i>$, where $|h^{\prime }1>=\alpha
_{1\uparrow }^{\dagger }|->$, $|h^{\prime }2>=\alpha _{1\downarrow
}^{\dagger }|\Theta >$, $|h^{\prime }3>=\frac{1}{\sqrt{2}}\alpha _{1\uparrow
}^{\dagger }\alpha _{1\downarrow }^{\dagger }(d_{1\downarrow }^{\dagger
}-d_{2\downarrow }^{\dagger })|0>$ , $|h^{\prime }4>=\frac{1}{\sqrt{2}}%
\alpha _{1\downarrow }^{\dagger }(d_{1\uparrow }^{\dagger }d_{1\downarrow
}^{\dagger }-d_{2\uparrow }^{\dagger }d_{2\downarrow }^{\dagger })|0>$, and $%
|h^{\prime }5>=\frac{1}{\sqrt{2}}(d_{1\uparrow }^{\dagger }d_{1\downarrow
}^{\dagger }d_{2\downarrow }^{\dagger }-d_{1\downarrow }^{\dagger
}d_{2\uparrow }^{\dagger }d_{2\downarrow }^{\dagger })|0>$, with the
corresponding eigenvalue $\lambda _{3,-1/2}$, obtained from the previous
matrix, changing $\Gamma $ by $-\Gamma $. From these two states, we obtain
the four particles states for $H_{ZBW}$ by adding one electron in the
decoupled $\alpha _{2\uparrow }$ state (the ground state of $H_{1}$)and we
have $\alpha _{2\uparrow }^{\dagger }|3,+1/2>_{0}$ with $S_{z}=+1$ and the
corresponding ground state energy $\lambda _{S_{z}=+1}=\varepsilon
_{F}-\Gamma +\lambda _{3,+1/2}.$ The first excited state corresponding to $%
S_{z}=0$ is given by $\alpha _{2\uparrow }^{\dagger }|3,-1/2>_{0}$ with $%
\lambda _{S_{z}=0}=\varepsilon _{F}-\Gamma +\lambda _{3,-1/2}.$ Therefore,
the lowest energy difference gives $E_{K}=\lambda _{S_{z}=+1}-\lambda
_{S_{z}=0}=\lambda _{3,+1/2}-\lambda _{3,-1/2}<0$, and we can not identify
this energy with a Kondo excitation because in the process there is no
spin-flip excitation ($\alpha _{2\downarrow }$ is absent in both states).
For $\Gamma =0$ , the model Hamiltonian gives $\lambda _{3,+1/2}=\lambda
_{3,-1/2}=\lambda _{3}$. For very large Coulomb repulsion ($U/|\Delta
|\rightarrow \infty $), Eq. (11) reduces to 3x3 matrix and we can solve to
obtain $\lambda _{3}=3(\varepsilon _{F}+\varepsilon _{d})/2-R,$ with $R=%
\sqrt{\Delta ^{2}+6V^{2}}$ and the corresponding eigenvector ($b_{1},b_{2}$,
and $b_{3}$) gives: $b_{1}=-\frac{1}{\sqrt{3}}\sqrt{1-\Delta /R\text{ }}$, $%
b_{2}=\frac{1}{\sqrt{6}}\sqrt{1-\Delta /R\text{ }}$, and $b_{3}=\frac{1}{%
\sqrt{2}}\sqrt{1+\Delta /R\text{ }}$. For small values of $\Gamma $ ($\Gamma
<<V$ and $\varepsilon _{F}=0$), we can write from Eq.(11), $\lambda
_{3,+1/2}\simeq 3\Delta -R-\Gamma V^{2}/(R^{2}+R\Delta )$ and $\lambda
_{3,-1/2}\simeq 3\Delta -R+\Gamma V^{2}/(R^{2}+R\Delta )$. So that, $%
E_{K}=-2\Gamma V^{2}/(R^{2}+R\Delta )$. For $V<<\Gamma $ and $\Gamma
<<|\varepsilon _{d}|$, we can write $\lambda _{3,+1/2}\simeq 2\varepsilon
_{d}-\Gamma -4V^{2}/(\Gamma -\varepsilon _{d})$, $\lambda _{3,-1/2}\simeq
2\varepsilon _{d}-\Gamma -2V^{2}/(\Gamma -\varepsilon _{d})$, and therefore $%
E_{K}=-2V^{2}/(\Gamma -\varepsilon _{d})$. From the above considerations,
for $\Gamma \neq 0$ and large values of $U$, we can see that $E_{k}<0$ for $%
\phi =0$ and $E_{k}>0$ for $\phi =\pi /2$. Therefore, there is always a
particular value $\phi =\phi _{c}$, with $0<\phi _{c}<\pi /2$, where $E_{k}=0
$.

B) For small Coulomb repulsion, in the I.V. regime ($U \lesssim |\varepsilon
_{d}-\varepsilon _{F}|$) and small values of $\Gamma/V$ (see Fig.~\ref{fig1}%
(c)), we can see that the ground state corresponds to $S_{z}=0 $ ($E_{k}>0 $%
). We can obtain this state solving $H_{0}$ in the four-particles subspace
with $S_{z}=0 $. So that, we obtain the ground state energy of $H_{ZBW}$ and
the corresponding eigenvector ($c_{1},c_{2},c_{3},c_{4}$, and $c_{5} $) from
the lowest eigenvalue $\lambda^{\prime }_{4,0}$ of the 5x5 matrix given by

\begin{equation}
\hspace{10mm}\left\vert 
\begin{array}{ccccc}
\varepsilon _{1}^{\prime } & -2V & 2V & 0 & 0 \\ 
-2V & \varepsilon _{2}^{\prime }-\Gamma  & 0 & -\sqrt{2}V & \sqrt{2}V \\ 
2V & 0 & \varepsilon _{2}^{\prime }+\Gamma  & \sqrt{2}V & -\sqrt{2}V \\ 
0 & -\sqrt{2}V & \sqrt{2}V & \varepsilon _{3}^{\prime } & 0 \\ 
0 & \sqrt{2}V & -\sqrt{2}V & 0 & \varepsilon _{4}^{\prime }%
\end{array}%
\right\vert ,
\end{equation}%
with $\varepsilon _{1}^{\prime }=2(2\varepsilon _{d}+U)$, $\varepsilon
_{2}^{\prime }=3\varepsilon _{d}+U+\varepsilon _{F}$, $\varepsilon
_{3}^{\prime }=2\varepsilon _{d}+U+2\varepsilon _{F}$, and $\varepsilon
_{4}^{\prime }=2\varepsilon _{d}+2\varepsilon _{F}$. The ground state reads: 
$|4,0>_{0}=\sum_{i}c_{i}|fi>$, where we define: $|f1>=d_{1\uparrow
}^{\dagger }d_{1\downarrow }^{\dagger }d_{2\uparrow }^{\dagger
}d_{2\downarrow }^{\dagger }|0>$, $|f2>=\frac{1}{\sqrt{2}}\alpha
_{1\downarrow }^{\dagger }(d_{1\uparrow }^{\dagger }d_{2\uparrow }^{\dagger
}d_{2\downarrow }^{\dagger }+d_{2\uparrow }^{\dagger }d_{1\uparrow
}^{\dagger }d_{1\downarrow }^{\dagger })|0>$, $|f3>=\frac{1}{\sqrt{2}}\alpha
_{1\uparrow }^{\dagger }(d_{1\downarrow }^{\dagger }d_{2\uparrow }^{\dagger
}d_{2\downarrow }^{\dagger }+d_{2\downarrow }^{\dagger }d_{1\uparrow
}^{\dagger }d_{1\downarrow }^{\dagger })|0>$, $|f4>=\frac{1}{\sqrt{2}}\alpha
_{1\uparrow }^{\dagger }\alpha _{1\downarrow }^{\dagger }(d_{1\uparrow
}^{\dagger }d_{1\downarrow }^{\dagger }+d_{2\uparrow }^{\dagger
}d_{2\downarrow }^{\dagger })|0>$, and $|f5>=\frac{1}{\sqrt{2}}\alpha
_{1\uparrow }^{\dagger }\alpha _{1\downarrow }^{\dagger }(d_{1\uparrow
}^{\dagger }d_{2\downarrow }^{\dagger }-d_{1\downarrow }^{\dagger
}d_{2\uparrow }^{\dagger })|0>$. The last term shows the antiferromagnetic
state $|\Omega >$for the impurities in this ground state. When this limit
take place, we can see (Fig.~\ref{fig1}(c))that always $E_{k}>0$ and the
fundamental state has $S_{z}=0$ for any value of $\phi $.

In Fig. 2 we show the zero temperature magnetic correlations $<\mathbf{S}%
_{1} \mathbf{S}_{2}>$ between the impurities ($\mathbf{S}_{1}$ and $\mathbf{S%
}_{2}$ are the spin $\frac{1}{2}$ operators impurities) as a function of $\phi $, for the same parameters of Fig.1.

\begin{figure}[ht]
\begin{center}
\includegraphics[width=6cm]{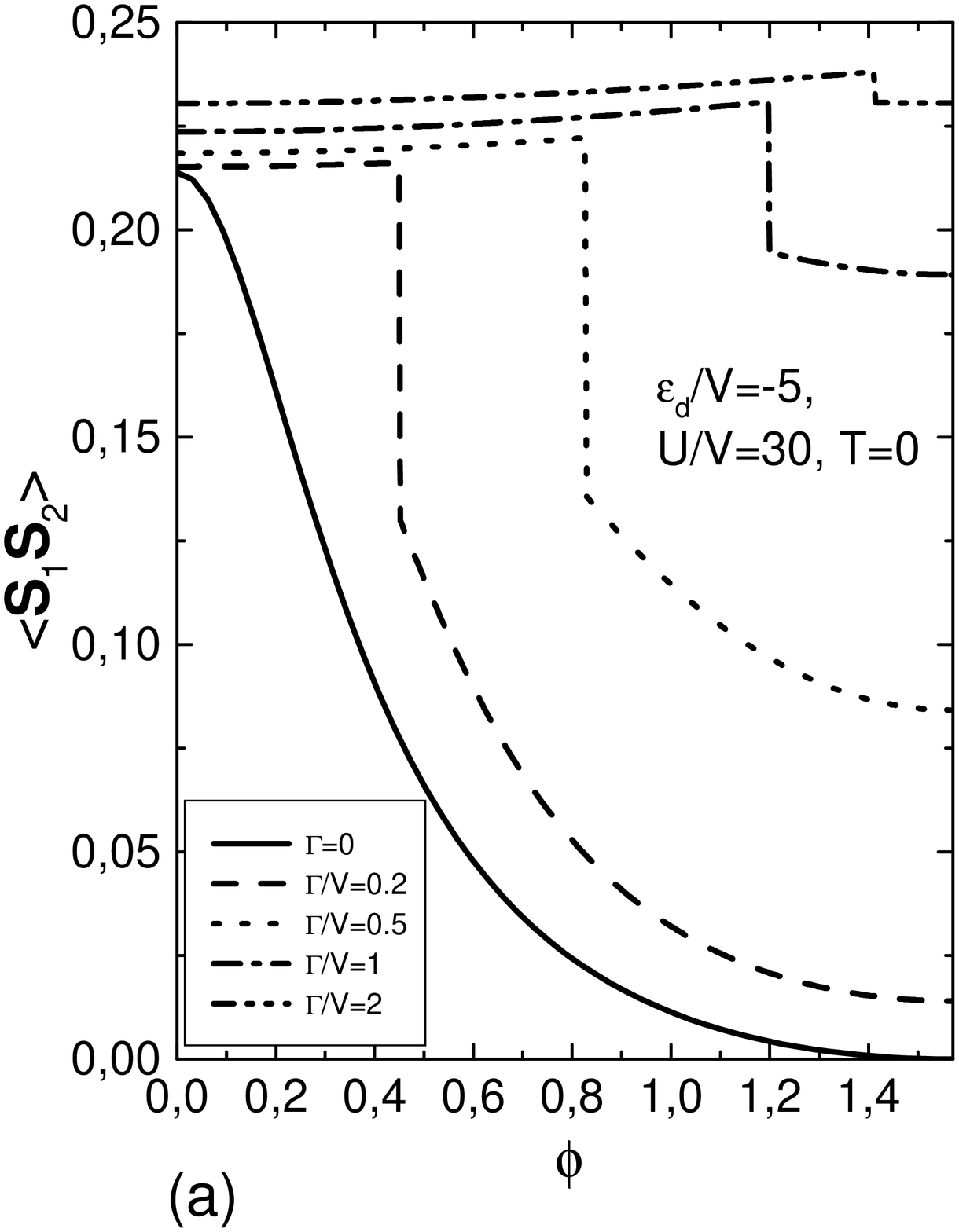} \includegraphics[width=6cm]{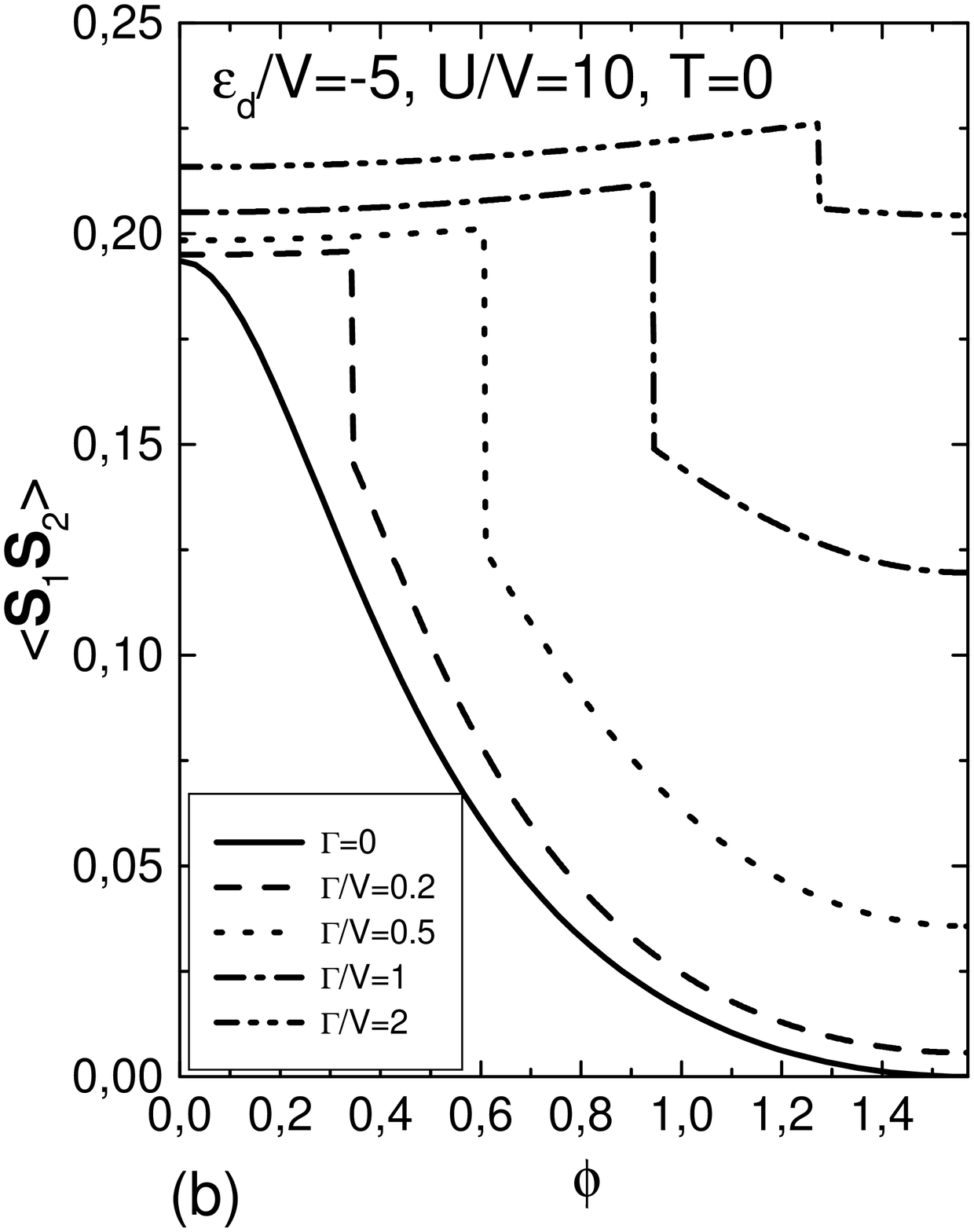}
\includegraphics[width=6cm]{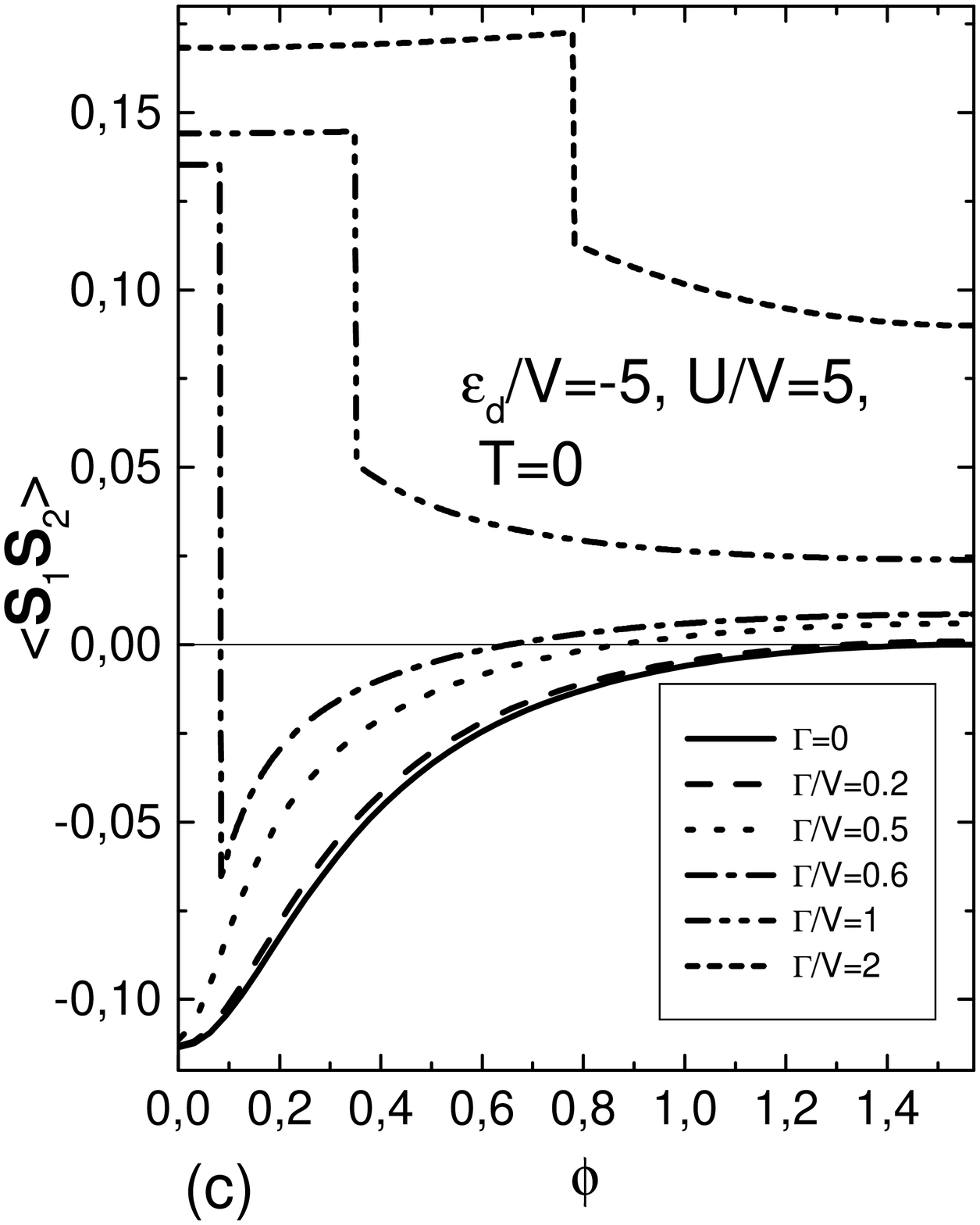}
\end{center}
\caption{Zero-temperature magnetic correlation $<\mathbf{S}_{1}\mathbf{S}_{2}>$ as a function of $\protect\phi $, for $\protect\varepsilon _{d}/V=-5$
and three different values of $U/V= $ 30 (a), 10 (b), and 5 (c).In (a) and (b)we take $\Gamma/V= $ 0, 0.2, 0.5, 1, and 2. For (c) we use $\Gamma/V= $ 0, 0.2, 0.5, 0.6, and 1 }
\label{fig2}
\end{figure}

For $\phi \rightarrow 0$ and large Coulomb repulsion ($|\varepsilon
_{d}-\varepsilon _{F}|<U$) we always observe ferromagnetic correlations
between the impurities (Fig.~\ref{fig2}(a)and (b)).

For $\phi =0$, we have
the ground state $\alpha _{2\uparrow }^{\dagger }|3,+1/2>_{0}$ and we can
write $<\mathbf{S}_{1}\mathbf{S}_{2}>=0.25(b_{1}^{2}+b_{2}^{2})$. When $\Gamma $ is increases, we can see that $|b_{1}|$ increases (see Eq. 11)
given more influence of the ferromagnetic state ($|h1>=\alpha _{1\downarrow}^{\dagger }|+>$) for the impurities in the ground state. As a consequence,
we can see that the ferromagnetic correlation increases with $\Gamma $ for $\phi =0$ at zero temperature. For $U/|\Delta |\rightarrow \infty $ and $\Gamma =0$, this ferromagnetic correlation reduces to $(1-\Delta /R)/8$.
When $\phi $ increases up to $\phi _{c}$ we can observe a 
%TCIMACRO{\U{a8}}%
%BeginExpansion
\"{}%
%EndExpansion
jump%
%TCIMACRO{\U{a8} }%
%BeginExpansion
\"{}
%EndExpansion
or discontinuity in $<\mathbf{S}_{1}\mathbf{S}_{2}>$ showing the transition
from $S_{z}=+1$ ground state to $S_{z}=0$. For $\phi =\pi /2$, the magnetic
correlation take the minimum value. This value, in the Kondo limit of each
impurity ($U/|\Delta |\rightarrow \infty $, $V/|\Delta |<<1$), gives $<%
\mathbf{S}_{1}\mathbf{S}_{2}>=0.25(a_{2}^{2}+\frac{2}{3}a_{3}^{2}-\frac{2}{%
\sqrt{3}}a_{1}a_{3})$ . For $\Gamma =0$, $a_{2}=a_{3}=0$, and $<\mathbf{S}%
_{1}\mathbf{S}_{2}>=0$. In (Fig.~\ref{fig2}(c)) we show the I.V. regime ($%
U\leq |\varepsilon _{d}-\varepsilon _{F}|$). For large values of $\Gamma /V$%
, we have the $S_{z}=+1$ ground state at $\phi =0$ and we can see that $<%
\mathbf{S}_{1}\mathbf{S}_{2}>$ has the same behavior observed in Fig.~\ref%
{fig2}(a) and (b). For small values of $\Gamma /V$, the $S_{z}=0$ ground
state take place and using $|4,0>_{0}$ we can write $<\mathbf{S}_{1}\mathbf{S%
}_{2}>=\frac{-3}{4}c_{5}^{2}$. So that, we have always antiferromagnetic
correlation between the impurities. Finally, for intermediate values of $%
\Gamma /V$ ($0.6$), we observe the transition from ferromagnetic to
antiferromagnetic correlation at $\phi =\phi _{c}$. In Fig.~\ref{fig1}(c),
for $\Gamma /V=0.6$ and $\phi =0$, we can see that a very small value of $|E_{K}|$ occurs. Due to this fact, the transition can take place only at
small values of $\phi _{c}$.

\begin{figure}[ht]
\begin{center}
\includegraphics[width=6cm]{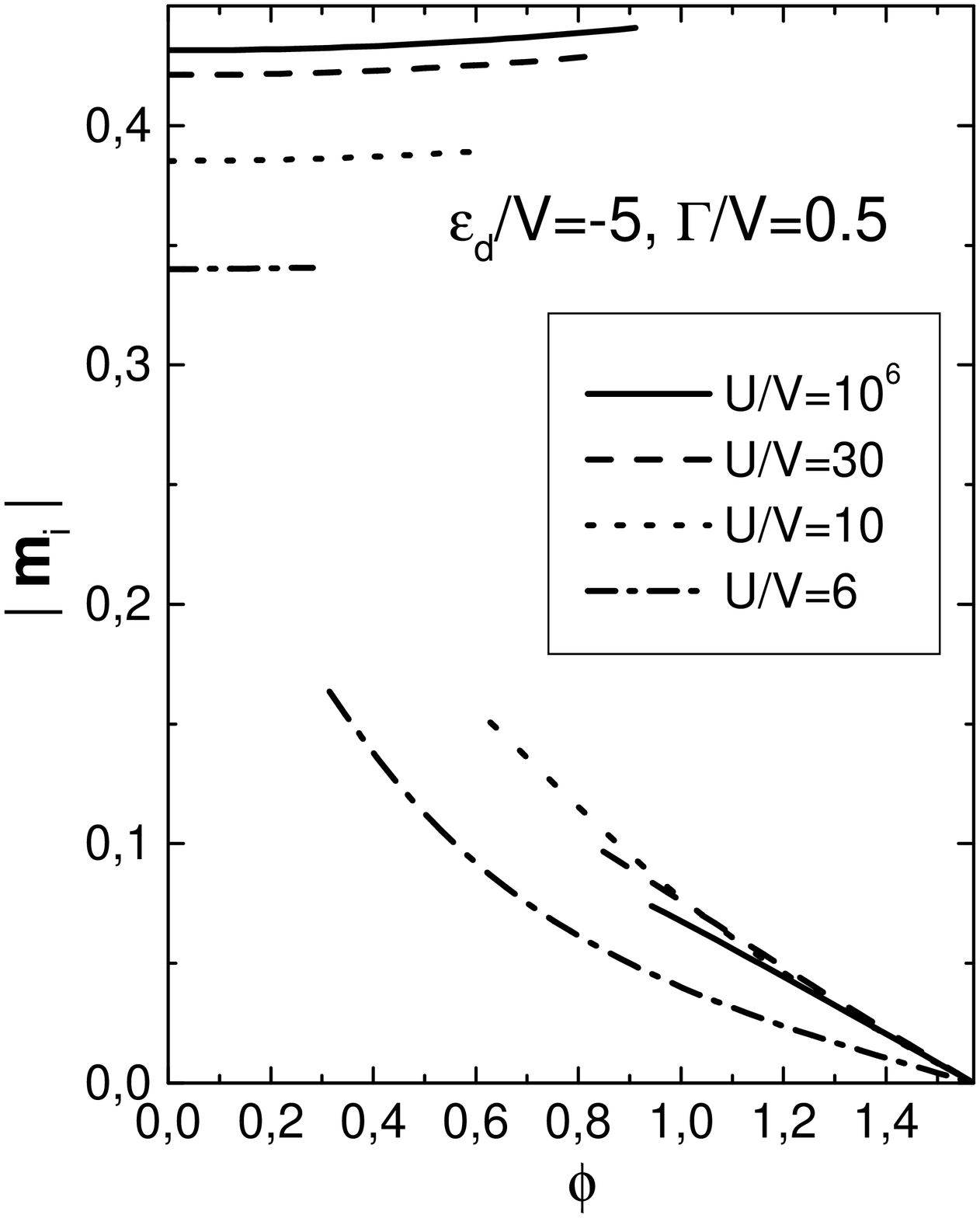}
\end{center}
\caption{The magnitude of the magnetic moment at each impurity site $|%
\mathbf{m}_{i}| $ as a function of $\protect\phi$, for $\protect\varepsilon _{d}/V=-5$, $\Gamma/V= 0.5 $, and four different values of $U/V= 10^{6}$,
30, 10, and 6.}
\label{fig3}
\end{figure}

For $\Gamma=0$ (solid lines in Fig.~\ref{fig1}and Fig.~\ref{fig2}), the
antiferromagnetic coupling between itinerant electrons disappears and the
model Hamiltonian is spin conserving. Therefore, the first triplet excited
state has the lower eigenvalue $\lambda _{S_{z}=1}$ (three times degenerate $%
S_{z}=\pm1,0$), and we can see that $E_{K}$ gives the low-energy spin
excitation in this model (Kondo energy). This energy decreases continuously
from the maximum value at $\phi =\pi/2 $, where two independent ($H_{Aj}$)
Anderson models take place, to zero for $\phi =0$, with the impurities in
the limit of very strong interaction regime, where the states $|3,\pm 1/2>$
play the role of an effective localized spin 1/2 which coupled to the band
states $\alpha _{2\sigma }^{\dagger }$ produce the physics that governs the
ground state of the Kondo model. Therefore, when the distance between the
impurities decreases, the interaction between the impurities via the
conduction electrons increases and reduces the $E_{K}$ energy. Furthermore,
in according to the Kondo Physics, the magnetic moment at each impurity site 
$j $ given by: $\mathbf{m}_{j}=\sum_{\sigma,\sigma^{\prime }}<d_{j\sigma
}^{\dagger }\mathbf{S}_{\sigma,\sigma^{\prime }}d_{j\sigma^{\prime }}> $,
where $\mathbf{S}_{\sigma,\sigma^{\prime }} $ are the standard Pauli
matrices, gives always zero for any value of $\phi$. On the contrary, for $%
\Gamma \neq 0$, the model Hamiltonian is spin non-conserving and we obtain $%
\mathbf{m}_{j}\neq 0$. We show in Fig.~\ref{fig3}, for $|\varepsilon
_{d}-\varepsilon _{F}|<U$, the magnitude of the magnetic moment $|\mathbf{m}%
_{j}| $ as a function of $\phi$.

The figure shows the region for $\phi < \phi_{c}$, where we observe a very
weak dependence on $\phi$ for the corresponding magnetic moment at $\phi =0$%
, where we can write $|\mathbf{m}_{j}|=0.25*(2b_{1}^{2}+b_{3}^{2}+b_{5}^{2}) 
$. For $\phi_{c}=\phi$, we observe the discontinuity showing the transition
from $S_{z}=+1$ ground state to $S_{z}=0$, and finally, for $\phi_{c}<\phi$
we can see that $|\mathbf{m}_{j}| $ decreases and reduces to zero at $\phi=
pi/2$ (see $|4,0>_{\pi /2} $). For $U/V= 6 $, far from the strong Kondo
limit, we can observe an important reduction of the magnetic moment. 
\begin{figure}[ht]
\begin{center}
\includegraphics[width=6cm]{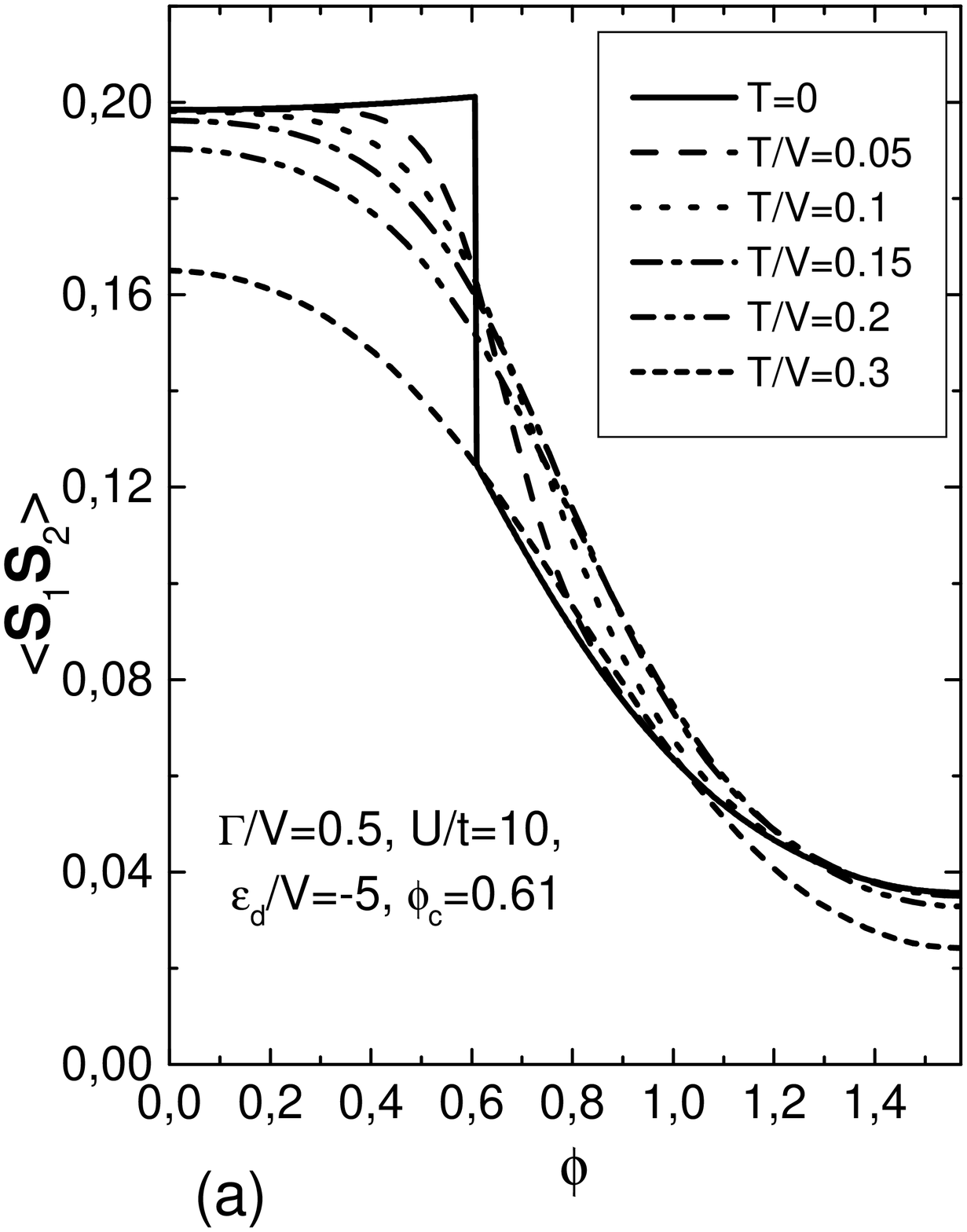} \includegraphics[width=6cm]{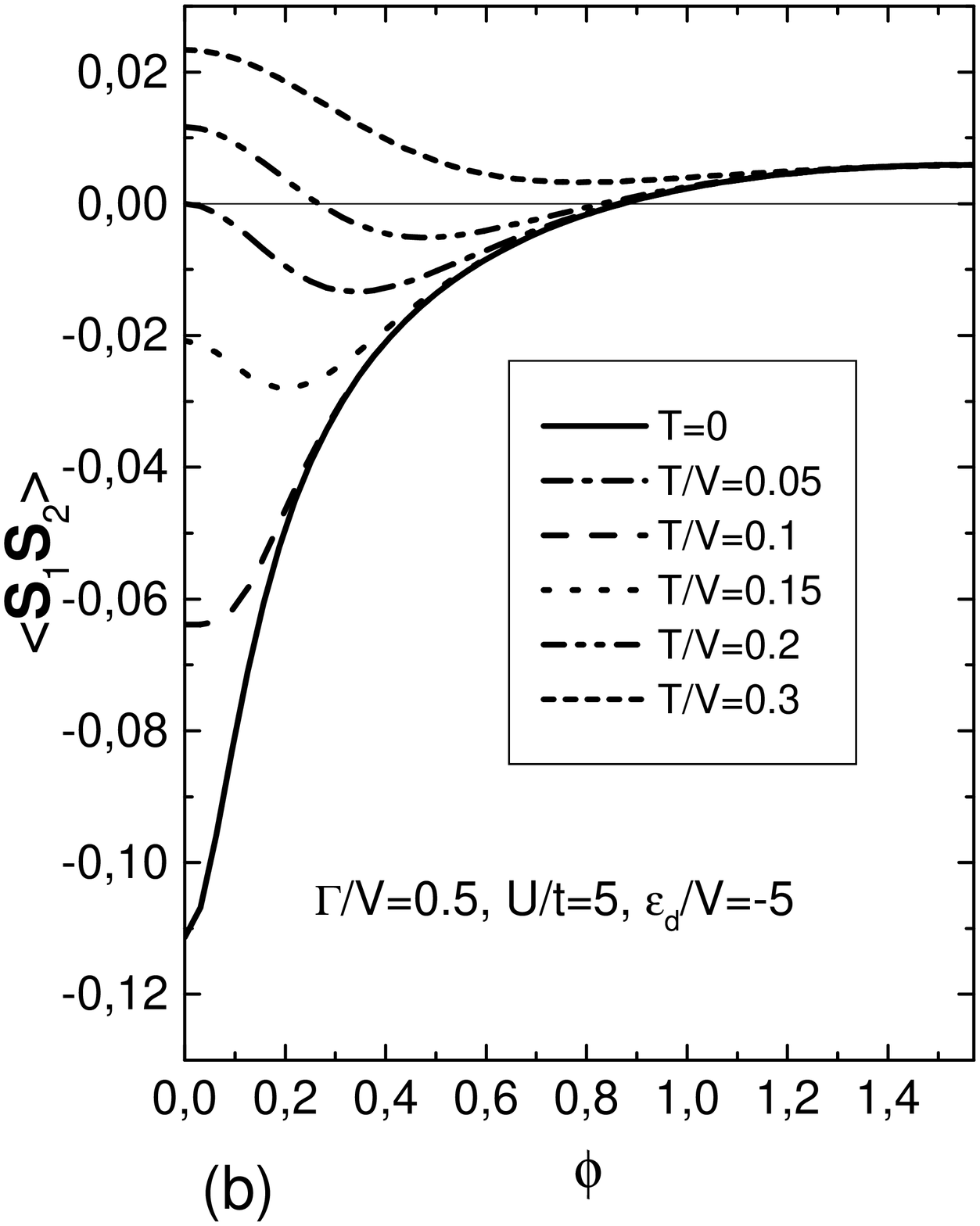}
\end{center}
\caption{Zero-temperature magnetic correlation $<\mathbf{S}_{1}\mathbf{S}%
_{2}>$ as a function of $\protect\phi $, for $\protect\varepsilon _{d}/V=-5$, $\Gamma/V=0.5 $ , and different values of temperature. In Fig.~\protect\ref{fig4}(a) we show the case of $U/V= 10$ and Fig.~\protect\ref{fig4}(b) shows 
$U/V= 5$ }
\label{fig4}
\end{figure}

In Fig. 4, we show the magnetic correlations $<\mathbf{S}_{1}\mathbf{S}_{2}>$
as a function of $\phi $, for $\varepsilon _{d}/V=-5$, $\Gamma /V=0.5$, and
different values of temperature ($T$). For $U/V=$ $10$, in the Kondo region,
Fig.~\ref{fig4}(a) shows different behavior depending on the value of $\phi $
related to $\phi _{c}$. For small values ($\phi <\phi _{c}\simeq 0.61$) and
very low temperatures, results show very strong ferromagnetic correlation
due to de ground state $\alpha _{2\uparrow }^{\dagger }|3,+1/2>_{0}$.
Therefore, as the temperature increases, the contribution of the low energy
levels reduce the magnetic correlation. On the contrary, for $\phi >\phi _{c}
$, it is interesting to note that at low temperatures, thermodynamical
excitations to the low excited states give additional contribution to the
ferromagnetic correlation. This is an expected result in a Kondo energy
level scheme (singlete-triplet structure). Therefore, we consider that $\phi
_{c}$ as the lowest limit of $\phi $ below which the breakdown of the Kondo
theory occurs. Finally, for $\phi \rightarrow \pi /2$, the splitting of the
low energy levels decrease, so that, correlation decreases with increasing
temperature.

For $U/V=$ $5$, in the I.V. regime, Fig.~\ref{fig4}(b) shows the
antiferromagnetic correlation between the impurities. We can see that $<%
\mathbf{S}_{1}\mathbf{S}_{2}>$ increases when $T $ increases. In Fig.~\ref%
{fig5}, we show the temperature dependence of $<\mathbf{S}_{1}\mathbf{S}%
_{2}> $ for $U/V=10$, $\varepsilon _{d}/V=-5$, $\Gamma/V= 0.5 $, and
different values of $\phi$, around the $\phi_{c}=0.61$.

\begin{figure}[ht]
\begin{center}
\includegraphics[width=6cm]{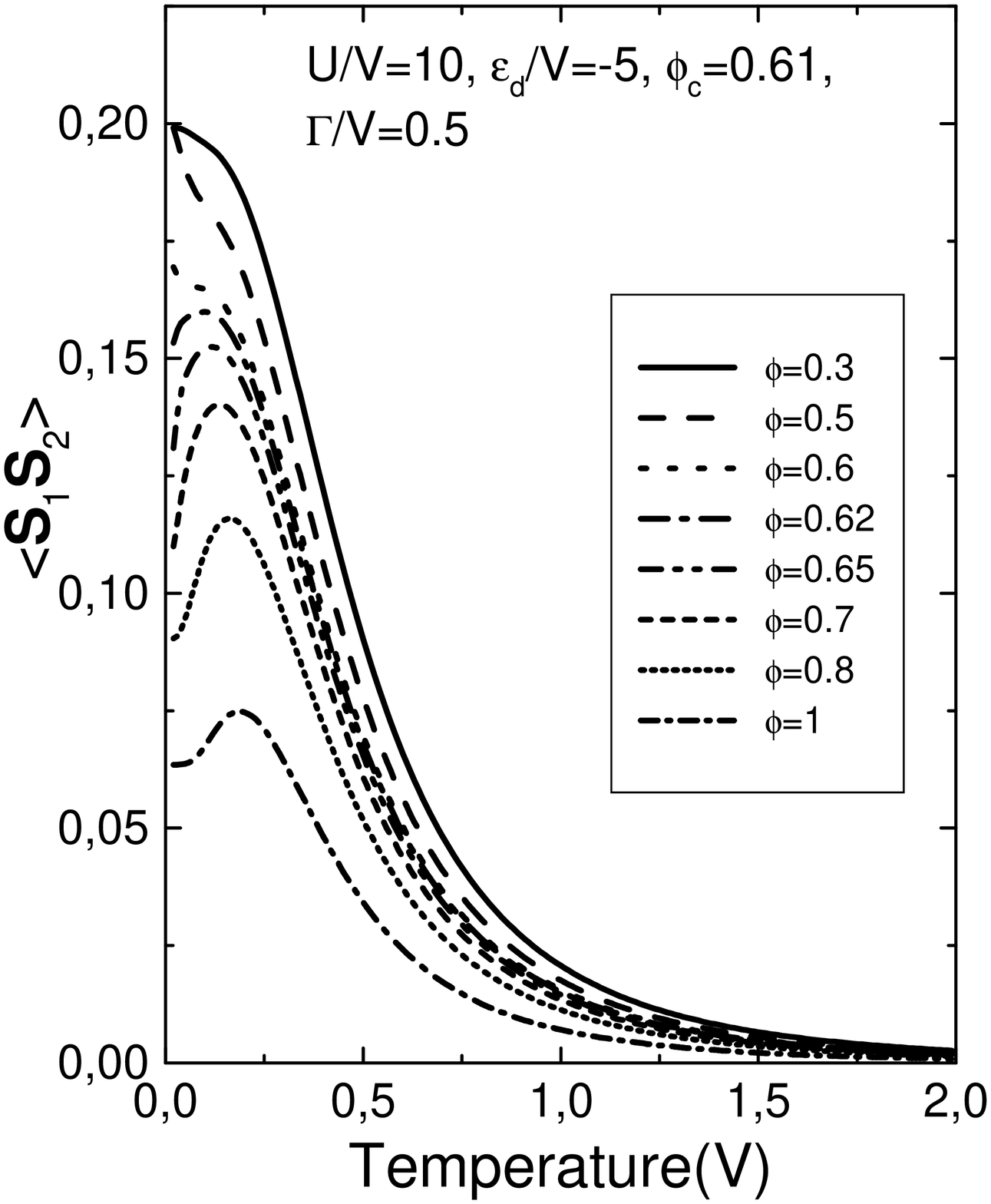}
\end{center}
\caption{the magnetic correlations $<\mathbf{S}_{1}\mathbf{S}_{2}>$ as a
function of temperature, for $\protect\varepsilon _{d}/V=-5$, $U/V=10$, $%
\Gamma/V= 0.5 $, and different values of $\protect\phi$ around the $\protect%
\phi_{c}=0.61$.}
\label{fig5}
\end{figure}

At low temperatures, for $\phi \geq \phi_{c}$ the curves show a maximum.
This maximum is due to the excitation from the $S_{z}=0 $ ground state to
the low excited states $S_{z}=+1, 0, -1$. When $\phi $ is increases from $%
\phi_{c} $, the maximum becomes more significant and according to the above
discussion in Fig.~\ref{fig4}, we consider that the temperature at the
maximum gives a rough measure of the Kondo temperature in this model. The
curves also show how the maximum moves to low temperatures (the Kondo
temperature goes down) when $\phi $ (the distance between impurities) is
decreased. For $\phi < \phi_{c}$, the maximum disappears and the Kondo
regime is impossible.

\section{CONCLUSIONS}

We have extended the zero-bandwidth limit of the two-impurity Anderson model
to include the effect of an antiferromagnetic gap in the conduction band
states. We have studied, as a function of $\phi =\mathbf{k}_{F}\mathbf{\cdot
r}$, the lowest excitation energy, the magnetic moment at each impurity
site, and the magnetic correlation between the impurities in this model. In
the region of parameters where the impurities are in the Kondo regime, as a
function of $\phi $, we have shown that a very interesting competition
between the AF gap and the Kondo physics of the two impurities take place.
At zero temperature, when the impurities are close enough ($\phi <\phi _{c}$%
), the AF splitting governs the physics of the system and the local moment
of the impurities are frozen in a state with very strong ferromagnetic
correlation between the impurities, and roughly independent of the distance.
On the contrary, when the impurities are sufficiently far apart ($\phi >\phi
_{c}$) and the AF gap is not too large, the scenario of Kondo physics takes
place: a non-magnetic ground state with the possibility of spin-flip
excitation can occurs. Here, the ferromagnetic $<\mathbf{S}_{1}\mathbf{S}%
_{2}>$ decreases when $\phi $ is increased from $\phi _{c}$, but the
complete decoupling of the impurities never occurs. In addition, the
presence of the AF gap gives a non-zero magnetic moment $\mathbf{m}_{j}$ at
each impurity site, showing a non complete kondo screening of the
impurities. Also, we can see that the residual magnetic moment decreases
when $\phi $ is increased. Finally, the zero-bandwidth limit approach used
here gives a new contribution to understand the very relevant and difficult
problem of two magnetic impurities in an antiferromagnetic metal. We expect
that new experimental results in nanodevices will confirm some of the
theoretical predictions obtained here.

\begin{center}
{\bf Acknowledgments}
\end{center}

The author acknowledges many illuminating discussions with Blas Alascio.This work was supported by the Consejo
Nacional de Investigaciones Cient\'{i}ficas y T\'{e}cnicas (CONICET).

{} 

\end{document}